\documentclass[aps,pra,reprint,amsmath,amssymb,floatfix,superscriptaddress]{revtex4-1}

\usepackage{physics}
\usepackage{color}
\usepackage{graphicx}
\usepackage{times,psfrag}
\usepackage{dsfont}
\usepackage{dcolumn}
\usepackage{bm,bbm}
\usepackage{latexsym,amsmath,amssymb,bm,euscript}
\usepackage[normalem]{ulem}
\usepackage[hidelinks]{hyperref}
\usepackage{microtype}
\usepackage{verbatim} % for comments

\newcommand{\MoSe}{MoSe$_{2}$}
\newcommand{\WS}{WS$_{2}$}
\newcommand{\MoS}{MoS$_{2}$}

\newcommand{\SiO}{SiO$_{2}$}
\newcommand{\SiN}{Si$_{3}$N$_{4}$}

\begin{document}
\title{Substrate interference and strain in the second harmonic generation from MoSe$_2$ monolayers}
% Force line breaks with \\
\author{S. Puri}
\affiliation{Department of Physics, University of Arkansas, Fayetteville, AR 72701, USA.}
\author{S. Patel}
\affiliation{Department of Physics, University of Arkansas, Fayetteville, AR 72701, USA.}
\author{J. L. Cabellos}
\affiliation{Universidad Polit\'ecnica de Tapachula. C.P.~30830. Tapachula, Chiapas, Mexico }
\author{L. E. Rosas-Hernandez}
\affiliation{Department of Physics, University of Arkansas, Fayetteville, AR 72701, USA.}
\author{S. Barraza-Lopez}
\affiliation{Department of Physics, University of Arkansas, Fayetteville, AR 72701, USA.}
\author{B. Mendoza}
\affiliation{Centro de Investigaciones en Optica, A.C.,  Le\'on, C.P.~37150. Guanajuato, Mexico}
\author{H. Nakamura}
\email{hnakamur@uark.edu}
\affiliation{Department of Physics, University of Arkansas, Fayetteville, AR 72701, USA.}

\begin{abstract}
Nonlinear optical materials of atomic thickness--such as non-centrosymmetric 2H transition metal dichalcogenide monolayers--have a second order nonlinear susceptibility ($\chi^{(2)}$) whose intensity can be tuned by strain. However, whether $\chi^{(2)}$ is enhanced or reduced by tensile strain is a subject of conflicting reports. Here, we grow high-quality MoSe$_2$ monolayers under controlled biaxial strain created by two different substrates, and study their linear and non-linear optical responses with a combination of experimental and theoretical approaches. A 15-fold overall enhancement in second harmonic generation (SHG) intensity is observed on MoSe$_2$ monolayers grown on SiO$_2$ when compared to its value when on a Si$_3$N$_4$ substrate. A seven-fold enhancement was ascertained to substrate interference, and a factor of two to the enhancement of $\chi^{(2)}$ arising from biaxial strain: substrate interference and strain are independent handles to engineer the SHG strength of non-centrosymmetric 2D materials.

\end{abstract}

\maketitle

Second harmonic generation (SHG) is a nonlinear optical phenomenon in which two photons combine and form a single photon with twice the energy of the original ones~\cite{Boyd2020}. SHG is used to determine crystal symmetry~\cite{malard2013observation,kumar2013second,shi20173r,klimmer2021all,hsu2014second} and it can also probe excitonic states~\cite{malard2013observation,kumar2013second,shi20173r,klimmer2021all,hsu2014second}.

Non-centrosymmetric monolayers (MLs) of transition metal dichalcogenides (TMDs) have very high nonlinear susceptibilities of the order of 10$^{5}$ - 10$^{8}$ pm$^2$/V, making them highly efficient for SHG~\cite{kumar2013second,malard2013observation,wen2019nonlinear}.
Furthermore, they can sustain large--uniaxial or biaxial--tensile strain~\cite{roldan2015strain,naumis2023mechanical}. 
Their crystal symmetry is lowered when under uniaxial strain, resulting in an increased angular anisotropy in polarization-resolved SHG measurements~\cite{mennel2019second,mennel2018optical,liang2017monitoring,li2023uniaxial,khan2020direct,mennel2020band} and a decrease in SHG intensity along the armchair direction~\cite{liang2017monitoring,mennel2020band}.

As it preserves the point symmetry of the crystal, {\em biaxial strain} could be considered a more desirable tool to tune the efficiency of nonlinearity without impacting polarization-dependent properties. Nevertheless, reports on SHG in biaxially tensile strained TMD MLs are scarce and contradictory. For instance, Covre \textit{et al}.~stated a {\em decrease} in the SHG intensity~\cite{covre2022revealing} on \MoSe\ MLs, while Liu \textit{et al}.~documented an {\em enhancement} in SHG in a \MoS\ ML~\cite{liu2023extraordinary}.
Similarly, Shi \textit{et al}.~observed an increased SHG in atomically thin \WS\ produced on silicon holes, which was explained as a result of a Fabry-Perot cavity resonance~\cite{shi2022giant}. Adding to the conflicting nature of some reports, the tuning rate for uniaxial strain in Reference~\cite{liang2017monitoring} was used to rule out the involvement of strain, even though the experimental geometry appears to more closely match one in which the 2D material is under biaxial strain~\cite{shi2022giant}.

\begin{figure*}
\begin{center}
\includegraphics[width=0.96\textwidth]{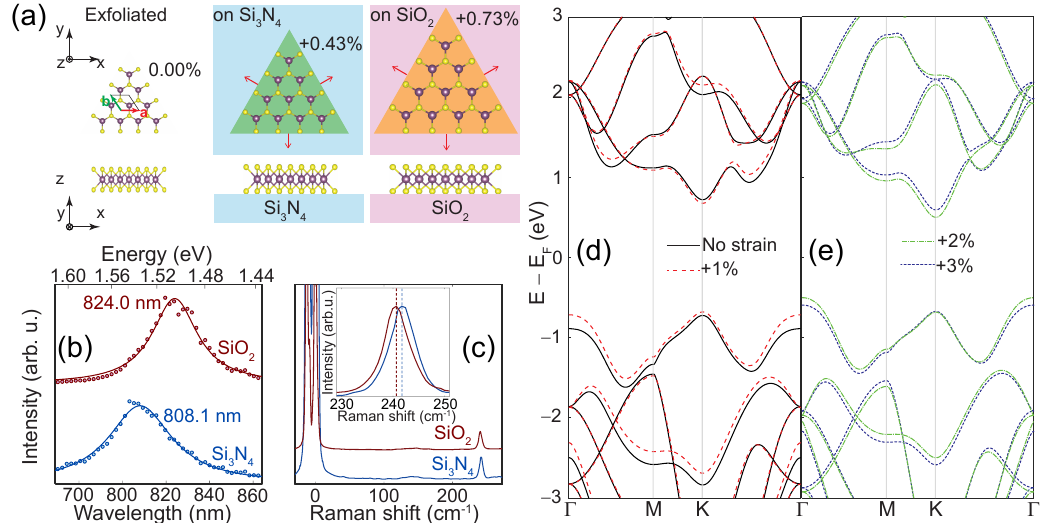}
\end{center}
\caption{(a) Schematics of MoSe$_2$ MLs without strain and under biaxial tensile strain by a rapidly quenched growth on Si$_3$N$_4$ or SiO$_2$ substrates. The disposition of the unit cell employed in SHG calculations and cartesian axes are depicted in the leftmost diagrams.
(b) PL of \MoSe\ MLs grown on \SiN\ and \SiO. Experimental data (dots) with Lorentzian fits (solid lines) show red-shift of the exciton peak on \SiO\ from biaxial tensile strain. (c) Raman shift in \MoSe\ MLs on \SiN\ and \SiO\ substrates. The inset highlights the $A_{1g}$ phonon peak for the MLs on each substrate, which is known to redshift for biaxial strain \cite{Horzum2013} but not for uniaxial strain. (d) PBE-DFT band structures of a MoSe$_2$ ML under no strain (solid lines) or 1\% tensile strain; the direct nature of the band gap (at the $K-$point) is apparent. (e) Electronic band structure under a 2\% and 3\% tensile strain.\label{fig:Fig1}
}
\end{figure*}

To resolve this puzzle, biaxially strained \MoSe\ MLs were grown by physical vapor deposition on either \SiN\ or \SiO\ substrates here. Strain was induced by a rapid thermal quenching after growth. The biaxial nature of strain as well as its magnitude were determined by optical characterization that included photoluminescence (PL), Raman studies [Figures \ref{fig:Fig1}(b) and \ref{fig:Fig1}(c)], and theory. Details follow.

A high growth temperature in the PVD process efficiently induces a tensile strain due to the thermal expansion mismatch between the ML and the substrate. We used two substrates: 73 nm tick \SiN\ on Si and 87 nm tick \SiO\ on Si. The percent of strain induced due to the thermal expansion mismatch is given by \cite{ahn2017strain}:
\begin{equation}
\mathrm{strain(\%)} = \left[{\int_{\mathrm{RT}}^{T_\mathrm{g}} (\alpha_\mathrm{MoSe_{2}}-\alpha_\mathrm{subs}})dT\right] \times 100,
\end{equation}
where $\alpha_\mathrm{MoSe_2}=7.1\times10^{-6}$  is the thermal expansion coefficient of bulk \MoSe\ \cite{el1976thermal} (which we assumed to be thickness-independent here).  $\alpha_\mathrm{subs}$ can either be $\alpha_\mathrm{Si_3N_4}=3.5\times10^{-6}$ \cite{sinha1978thermal}, or $\alpha_\mathrm{SiO_2}=5.0\times10^{-7}$ \cite{roy1989very}. $T_\mathrm{g}$ is the growth temperature and $\mathrm{RT}$ stands for room temperature. Additional details will be presented elsewhere (S. Patel, manuscript in preparation). Our estimates yield up to 0.43\%, or 0.73\%  tensile strain at $T_\mathrm{g}=1130^\circ$ C when the substrate is \SiN\ or \SiO, respectively. (See Figure \ref{fig:Fig1}(a) for schematics, and Methods for details.)

When contrasted with a peeled, unstrained reference MoSe$_2$ ML sample, the PL shows a redshift of the exciton peak for samples grown on substrates, which is a signature of biaxial tensile strain. In turn, the Raman shift measured for \MoSe\ MLs on both \SiN\ and \SiO\ substrates also shows a redshifted $A_{1g}$ phonon peak (known to be significant only under biaxial strain \cite{Horzum2013}). As depicted in Figure \ref{fig:Fig1}(d), MoSe$_2$ MLs display direct bandgap at the $K-$point when under no strain, which reduces in size under a 1\% biaxial tensile strain. As seen in Figure \ref{fig:Fig1}(e), further (2 and 3\%) tensile strain induces indirect bandgap (in between the $\Gamma$ and $K$ points) \cite{Ugeda} (see Methods).

\begin{figure*}
\begin{center}
\includegraphics[width=0.96\textwidth]{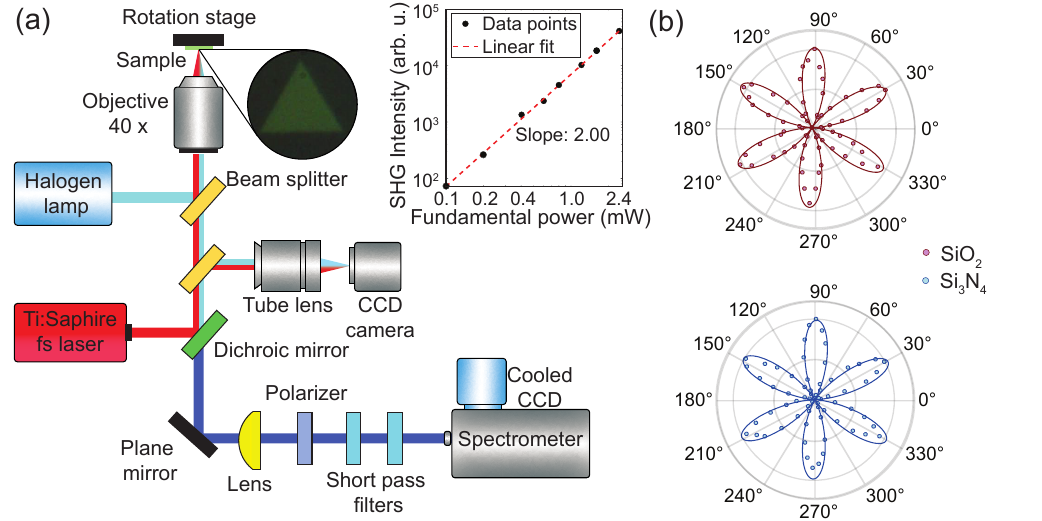}
\end{center}
\caption{(a) Sketch of the experimental setup for the collection of SHG light in reflection geometry using a tunable femtosecond laser. The inset shown is double logarithmic plot of pump power dependence of SHG. The slope of the plot is 2. (b) {\em Normalized} SHG intensity as a function of sample rotation angle \(\theta\) about the $z-$axis, in which the polarization of the SHG signal is parallel to that of the pump laser. Experimental data points are represented by pink dots for \MoSe\ MLs on \SiO\, and by blue dots for \MoSe\ MLs on \SiN, and the solid line is calculated theoretically to be proportional to $\sin^{2}(3 \theta)$.\label{fig:Fig2}
}
\end{figure*}

We now turn to SHG measured by a setup schematically shown in Figure \ref{fig:Fig2}(a). The excitation source is a Ti:Sapphire femtosecond laser  (Tsunami, Spectra-Physics) generating 200 fs pulses at a repetition rate of 80 MHz (see Methods). An inset in Figure \ref{fig:Fig2}(a) shows a log-log plot of pump power {\em versus} SHG intensity of the \MoSe\ ML with a slope of 2, which is a hallmark of SHG. For polarization-resolved SHG measurements, the sample is rotated in 5$^\circ$ increments, and a linear polarizer is used in the detection line to collect SHG signals with a polarization parallel to that of the pump laser. The polarization resolved SHG depicted in Figure \ref{fig:Fig2}(b) shows a characteristic sixfold symmetry that arises from the $D_{3h}$ symmetry of the underlying crystal\cite{mennel2018optical,malard2013observation,kim2015optical}, {\em even under biaxial strain}.

%%%%%%%%%%%%%%%%%%%%%%%%%%%%%%%%%%%%%%%%%%%%%%%%%%%%%%%%%%%%%%

Following Wang and Qian \cite{Qian}, we calculate the effective second order susceptibility for a parallel polarization configuration.
By expanding the expression for the SHG polarization:
\begin{equation}\label{eq:Pol}
P_{i,2\omega} = \epsilon_0 \sum_{j = x,y,z} \sum_{k = x,y,z} \chi_{ijk}^{(2)}(2\omega;\omega,\omega) E_{j}E_{k},
\end{equation}
Where $\epsilon_0$ is the permittivity. 
There is only one independent entry in the second-order dielectric susceptibility for point group $\bar{6}m2$ ($D_{3h}$)\cite{Boyd2020}: $\chi_{yyy}^{(2)}= - \chi_{yxx}^{(2)} = - \chi_{xxy}^{(2)} = - \chi_{xyx}^{(2)}\equiv \chi^{(2)}$, where the $x$, $y$ and $z$ labels correspond to the laboratory axes as shown in Figure \ref{fig:Fig1}(a) and we dropped the dependency on $2\omega$ and $\omega$ from $\chi^{(2)}$. Then the second harmonic polarization takes the form:
\begin{align}\label{eq:pol2}
P_{x,2\omega} &= -2\epsilon_0 \chi^{(2)} E_x E_y ,\\ 
P_{y,2\omega} &= \epsilon_0 \chi^{(2)} \left( E_y E_y - E_x E_x \right),\\
P_{z,2\omega} &= 0.
\end{align}
 In a polarization-resolved SHG, an analyzer is inserted to select SHG polarized either parallel or perpendicular to that of the pump beam, while rotating the polarization of the pump on the sample. The electric field of the pump at an angle of $\theta$ with respect to the zigzag ($x$) direction of \MoSe\ ML can be expressed as $\mathbf{E}=(E_0\cos\theta,E_0\sin\theta,0)$. The effective second order susceptibility in the parallel configuration $\chi_{||}^{(2)}$ (see Ref. \cite{Qian}) is obtained by applying a projection onto a vector parallel to the polarization of the incoming beam:
\begin{equation}
\chi_{||}^{(2)}(\theta) =  (\cos{\theta}, \sin{\theta},0)\cdot \frac{\mathbf{P}}{\epsilon_0 E_0^2},
\end{equation}
with $\mathbf{P}$ as in Equation (\ref{eq:pol2}). This way:
\begin{equation}
\chi_{||}^{(2)}(\theta) =   \chi^{(2)} (-3\cos^2{\theta}\sin{\theta}+\sin^{3}{\theta})=-\chi^{(2)}\sin(3\theta),
\end{equation}
and the SHG intensity in the polar plot [solid curves in Figure \ref{fig:Fig2}(b)] can be expressed as:
\begin{equation}
I_\mathrm{SHG}(\theta) \propto |\chi^{(2)}|^{2} \sin^{2}(3 \theta).
\end{equation}
The SHG is thus proportional to $|\chi^{(2)}|^{2}$, which will be later used as a measure of SHG intensity.

%%%%%%%%%%%%%%%%%%%%%%%%%%%%%%%%%%%%%%%%%%%%%%%%%%%%%%%%

\begin{figure*}
\begin{center}
\includegraphics[width=0.96\textwidth]{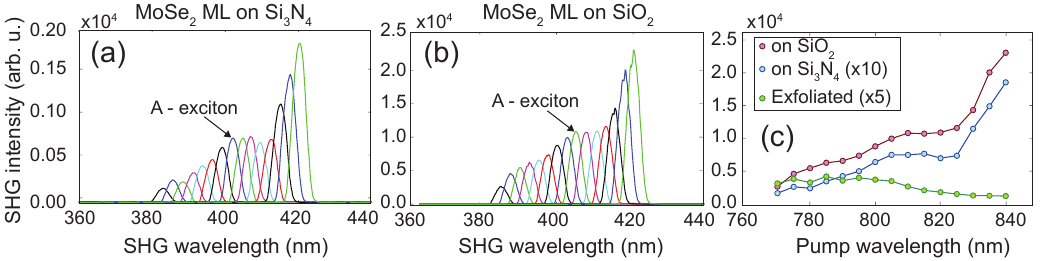}
\end{center}
\caption{(a) SHG intensity as a function of the SHG wavelength for the ML sample on \SiN. (b) SHG intensity as a function of the SHG wavelength for the ML sample and \SiO. The shoulders on the SHG spectrum in (a) and (b) are attributed to single photon resonance at A-exciton at around the pump wavelengths of 805 nm and 810 nm for \MoSe\ MLs on \SiN\ and \SiO, respectively, which is consistent with a different biaxial tensile strain being created by the two different substrates. (c) A 15-fold enhancement in the SHG response from the \MoSe\ ML on \SiO\ when contrasted to that on \SiN\ across the measured wavelengths.
\label{fig:Fig3}}
\end{figure*}

At this point, we address the main contribution from this work, {\em i.e.}, the {\em relative magnitude} of the SHG signals obtained on the two biaxially-strained samples. To this end, Figures \ref{fig:Fig3}(a) and \ref{fig:Fig3}(b) show the SHG intensity as a function of the SHG wavelength for the tensile strained ML samples grown on \SiN\ and \SiO{} substrates, respectively. The overall increase in the SHG intensity towards longer wavelength resembles earlier results \cite{kikuchi2019multiple,miyauchi2016influence}. We also observed a shoulder on the SHG spectrum, which is attributed to single photon resonances known as A-excitons at pump wavelengths of 805 nm and 810 nm for \MoSe\ MLs on \SiN\ and \SiO, respectively. The different shifts in A-exciton peaks confirm the different biaxial tensile strain on the two different substrates discussed in Figures \ref{fig:Fig1}(b) and \ref{fig:Fig1}(c).

A ratio of SHG intensity across the measured wavelength range is provided in Figure \ref{fig:Fig3}(c). The main takeaway is a large, {\em nearly 15-fold enhancement in the SHG response} of the \MoSe\ ML on 87 nm thick \SiO\ when compared to that of a \MoSe\ ML on 73 nm thick \SiN.  On the other hand, quantitatively comparing the SHG intensity of strained MoSe$_2$ MLs with the exfoliated ML poses a challenge because (1) intrinsic material properties (such as doping) may differ between the exfoliated {\em versus} PVD-grown \MoSe\ MLs, and (2) the 285 nm thick \SiO\ substrate used for the exfoliated ML happens to be close to a minimum of interference, which makes the interference contribution highly sensitive to a small error in the real thickness of the dielectric layer. In what follows, we focus on PVD grown \MoSe\ ML samples to determine the contribution of strain on the observed SHG enhancement, combining experimental data and computationally obtained SHG responses under strain.

%%%%%%%%%%%%%%%%%%%%

To calculate the effect of substrate interference on the SHG response, we use the model by Song {\em et al.}~\cite{song2023interference}. For the normally incident pump of wavelength $\lambda$ and intensity $I_{\lambda}$ on \MoSe\ ML on a dielectric film of thickness $d$ and a Si substrate, the SHG intensity $I_{\lambda/2}$ is given by\cite{song2023interference}:
\begin{equation}
        \ I_{\lambda/2} = \frac{1}{2\epsilon_0c}  |\beta_\lambda|^2  \frac{2\pi\chi^{(2)}}{\lambda \epsilon_0}  I_\lambda^2,
\end{equation}
where $\beta_\lambda$ is the structure factor that encapsulates the influences arising from the layered structure given as:
\begin{equation}
    \ \beta_\lambda = (1 + R_\lambda)^2 (1 + R_{\lambda/2}).
\end{equation}
Here $R_\lambda$ and $R_{\lambda/2}$ are the reflection coefficients of whole structure of $\lambda$ and $\lambda$/2 which are given by:
\begin{equation}
 \ R_\lambda = r \hspace{0.1cm} + \hspace{0.1cm} \frac{R_{s,\lambda}t^2}{1 - R_{s,\lambda}r}, \hspace{0.3cm}
    \ R_{s,\lambda} = \frac{r_{01} + r_{12} e^{2i\omega n_{1}d}}{1 + r_{01} r_{12} e^{2i\omega n_{1}d}}.
\end{equation}
In addition, $r = - \eta$/(1 + $\eta$) and $t = 1/(1 + \eta)$ are the reflection and transmission coefficients of a \MoSe\ ML, respectively, and $\eta$ is calculated from the effective bulk refractive index $n_{2D}$ as $\eta = -ih\omega (n_{2D}^2 - 1)/2$ with $\omega = 2\pi/\lambda$. The reflection coefficient at the interface between two layers is given by $r_{ij} = (n_i - n_j)/(n_i + n_j)$.

\begin{figure*}
\begin{center}
\includegraphics[width=0.96\textwidth]{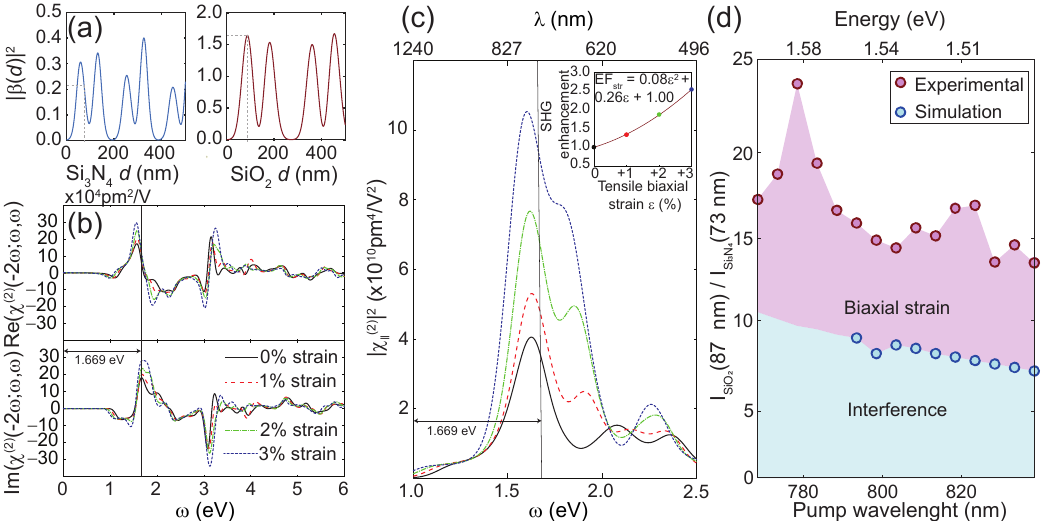}
\end{center}
\caption{(a) Substrate interference $\beta^2$ {\em versus} dielectric thickness $d$ for \SiN\ (left subplot) and \SiO\ (right subplot) slabs. The gray lines highlight the thickness of the dielectric films used in our experiments. (b) Real and imaginary contributions to $\chi^{(2)}(\omega)$. (c) Evolution of the SHG intensity due to biaxial strain alone, as a function of frequency. ($EF_{str}$ in the inset is the enhancement factor due to biaxial tensile strain for $\omega=1.669$ eV.) (d) Experimental observation of SHG relative enhancement, and elucidation of the separate effects of interference and biaxial strain.\label{fig:Fig4}
}
\end{figure*}

The enhancement factor due to interference ($\mathrm{EF_{int}}$) is thus written as:
\begin{equation}
  \mathrm{EF_{int}} = \left|{\frac{\beta_{\lambda}(SiO_{2})}{\beta_{\lambda} (Si_{3}N_{4})}}\right|^2,
\end{equation}
and displayed in Figure \ref{fig:Fig4}(a) for the two substrates used here.
%{\bf We could match $\lambda$ and use $\chi^2$ as calculated to redo Figs 4a and 4b.} Salva's comment??
The calculation assumes \MoSe\ ML on SiO$_2$($d$ nm)/Si or Si$_3$N$_4$($d$ nm)/Si substrates as a function of $d$ under an 800 nm pump laser. The delineated gray dashed lines serve as indicators for the enhancement resulting from interference, corresponding to 87 nm (73 nm) thick \SiO\ (\SiN) substrates utilized in our experiment. The results indicate that an $\mathrm{EF_{int}}$ as large as 7.2 could be attributable to constructive interference. We note that the interference could be overestimated by about 20-30\% due to the use of plane waves in the calculation, as the experiment focuses beam through the high NA objective lens \cite{song2023interference}.

%%%%%%%%%%%%%%%%%%%%

The real and imaginary parts of $\chi^{(2)}$ as a function of $\omega$ calculated with the TINIBA code (developed by J.~L.~Cabellos, T.~Rangel and B.~Mendoza) are provided in Figure \ref{fig:Fig4}(b). In Figure \ref{fig:Fig4}(c), we depict the energy dependency of $|\chi^{(2)}|^2$ (proportional to $I_\mathrm{SHG}$), which is seen to increase monotonically with applied tensile biaxial strain at excitation energies (1.48-1.61 eV) studied by the experiment (Figure \ref{fig:Fig4}(d)). Thus, the theory reproduces the tensile-strain induced {\em enhancement} of SHG by the experiment. Furthermore, both theory and experiment highlight the importance of resonant features that originate from band-to-band transitions, as seen by steeply changing SHG intensity as a function of wavelength. The theory also suggests that even qualitatively different strain effects may be expected in other wavelength ranges near the resonance, which may explain widely varying strain effects in the literature.

In summary, we explain two independent contributors to the enhancement of SHG in non-centrosymmetric crystals: biaxial tensile strain, and interference originating at the supporting substrate. This way, we provide clarity on two independent factors affecting the magnitude of the SHG on 2D non-centrosymmetric crystals unequivocally. These results shed light on the biaxial strain tuning of nonlinear optical process in atomically thin materials, and provide invaluable guidelines for future nanoscale nonlinear/quantum optical devices \cite{moody20222022,autere2018nonlinear}.

\section{Methods}

\subsection{Experimental}  \MoSe\ MLs were grown by physical vapor deposition (PVD) at 1130$^\circ$ C. We used \MoSe\ from AlfaAser (99.9 \% pure) as a precursor under argon flow. The diameter of a quartz tube was 1 inch. We employed reverse flow of argon from substrate towards the powder end to avoid any unwanted nucleation before the growth temperature is reached. Once the growth temperature is reached, forward flow is used for short period of time (typically 10s) for growth and then switched back to the reverse flow. Rapid quenching of the growth was carried out by sliding the quartz tube out of the hot zone of the furnace. Details on biaxial strain induced by high temperature PVD will be presented elsewhere (S. Patel, manuscript in preparation). As a comparison, we also measured exfoliated \MoSe\ MLs purchased from a vendor (2D semiconductors) on \SiO.

For the polarization resolved SHG setup, we used a Ti:Sapphire femtosecond (fs) laser source (Tsunami, Spectra-Physics) generating pulses of duration 200 fs at repetition rate of 80 MHz. With the use of a prism compensator, the pulse width was kept constant between 770 nm and 840 nm. A dichroic mirror directs excitation laser to 40x objective lens (NA 0.6) to excite the sample and the SHG signal is collected in the reflection geometry. A beam splitter added on the beam path illuminates the sample with white light and another beam splitter directs the reflected light to a CCD camera for sample imaging. Two short pass filters were used to reject the pump beam to detect SHG signals in a 500 mm focal spectrometer (Acton Research) equipped with a thermo-electrically cooled CCD. For the polarization resolved SHG measurements, sample is rotated in 5$^\circ$ step and a linear polarizer is used in the detection line to collect SHG signals parallel to the pump polarization.

\subsection{Theoretical}
First-principles calculations were carried out using plane wave density functional theory (DFT) as implemented in the ABINIT~\cite{Gonze2002} package. The General Gradient Approximation (GGA) with Perdew-Burke-Ernzerhof~\cite{PBE1996} (PBE) functional was considered, along with optimized norm-conserving Vanderbilt pseudopotentials~\cite{Hamann2017}. The plane-wave energy cutoff was set to 20 Hartrees ($~500$ eV) and a Monkhorst-Pack~\cite{Monkhorst1976} \textit{k}-point sampling of $18\times 18\times1$ centered in the $\Gamma$ point was used. The results indicate that the pristine \MoSe\ ML  has a lattice constant of $a = 3.327 $ \AA, a distance between selenium bonds of $d_{Se-Se} = 3.344 $ \AA, and a band gap of $E_g = 1.445 $ eV, which are in good agreement with previously reported values~\cite{Chang2013,Horzum2013}. After considering a biaxial tensile strain of $1\%$, $2\%$ and $3\%$ we found that at $1\%$ the band gap remains direct as in the pristine case but undergoes from direct to indirect as more strain is applied as predicted before~\cite{Horzum2013}. The second order susceptibility tensor was calculated with the TINIBA code, developed at the Centro de Investigaciones en Optica by B.~Mendoza, J.~L.~Cabellos, and T.~Rangel.

We used a scissor correction equal to 0.809 eV on the electronic band structure of all calculated susceptibilities; this value was taken as the GW gap reported in Reference \cite{Ugeda}. Our calculations are of the single-particle type, so the calculated responses do not include excitonic effects.

\section{Author Contributions}
HN conceived the project and directed all experimental work. S Puri performed the non-linear optical characterization and analyses. S Patel grew the MoSe$_2$ ML samples and performed PL and Raman measurements. SBL directed the theoretical work in which LER-H optimized the MoSe$_2$ MLs under strain and JLC carried out SHG calculations in a code developed by BM. LER-H crafted the final version of all figures. All authors discussed the results. S Puri, SBL, and HN wrote the manuscript with input from all the authors.

\section{Notes}
The authors declare no competing financial interest.

\begin{acknowledgements}
The nonlinear optics studies were supported by the Department of Defense under Award No. FA9550-23-1-0500. The growth system used to prepare MLs was supported by the MonArk NSF Quantum
Foundry from the National Science Foundation under Award No. DMR-1906383. L.E.R.-H. and S.B.-L. were funded by the US Department of Energy (Award No.~DE-SC0022120). Calculations performed at Arkansas were carried out at the Pinnacle Supercomputer, funded by the NSF under award OAC-2346752.
\end{acknowledgements}

\bibliographystyle{apsrev4-1} % Tell bibtex which bibliography style to use
\bibliography{MoSe2-SHG}

\end{document}